\documentclass[%
reprint,
 amsmath,amssymb,
 aps, physrev,
prb,
amsmath,amssymb,
 ]{revtex4-2}

\usepackage{graphicx}
\usepackage{dcolumn}
\usepackage{bm}
\usepackage{color}
\usepackage{xcolor}
\usepackage{rotating}
\usepackage{makecell}
\usepackage{placeins}
\usepackage{circledtext}
\usepackage[normalem]{ulem}
\usepackage{graphicx}
\usepackage{subcaption}
\usepackage{tikz}
\usepackage{cancel}

\renewcommand{\vec}[1]{\mathbf{\boldsymbol{#1}}}

\definecolor{mygreen}{RGB}{138, 182, 7}
\definecolor{mybrown}{RGB}{204, 153, 0}

\begin{document}

\preprint{APS/123-QED}

\title{\textbf{Optical properties of hydrogenated graphene superlattices} 
}%

\author{V.~A.~Saroka}
\email{vasil.saroka@roma2.infn.it}%
\affiliation{%
Department of Physics, University of Rome Tor Vergata and INFN, Via della Ricerca Scientifica 1, 00133 Roma, Italy
}%
\affiliation{Institute for Nuclear Problems, Belarusian State University, Bobruiskaya 11, 220006 Minsk, Belarus}

\author{O.~Pulci}
\email{Olivia.Pulci@roma2.infn.it}
\affiliation{%
Department of Physics, University of Rome Tor Vergata and INFN, Via della Ricerca Scientifica 1, 00133 Roma, Italy
}%

\author{M.~D'Alessandro}
\email{marco.dalessandro@ism.cnr.it}
\affiliation{%
 Istituto di Struttura della Materia-CNR (ISM-CNR),
 Via del Fosso del Cavaliere 100, 00133 Roma, Italia
}%




\date{\today}

\begin{abstract}

We investigate with first principles theoretical methods optical response accounting for excitonic effects of quasi-one dimensional monolithic arrays of graphene nanostripes --  two-dimensional graphene superlattices, where quasi-metallic and dielectric regions alternate by selective hydrogenation of graphene. It is shown that chemical engineering of the graphene surface can lead to strong and well-isolated absorption peaks in the far-infrared and possibly even terahertz frequencies. The band gap scaling suggests no excitonic insulators.
\end{abstract}

\maketitle


\section{\label{sec:Intro}Introduction}

Graphene is a monolayer of carbon atoms arranged in a honeycomb lattice, which is mechanically strong, flexible and chemically stable, enabling free-standing forms~\cite{Novoselov2004}. Its electronic spectrum features a gapless Dirac cone spectrum in the wide range of electronic energies~\cite{Novoselov2005}. The optical absorption spectrum of the Dirac cone yields a quantized constant $2.3\%$ absorption~\cite{Nair2008} that is also characteristic to other two-dimensional (2D) materials such as silicene, germanene and stanene~\cite{Matthes2013,Matthes2014} and even conventional 2D semiconductors~\cite{Fang2013,Merthe2016}.

The electronic applications of graphene requires opening a band gap in the electronic energy spectrum and a lot of efforts have been undertaken in this direction: carrier confinement~\cite{Tapaszto2008}, chemical modification~\cite{Withers2011}, substrate engineering~\cite{Zhou2007} etc. 
Optically wise, gap opening in the Dirac spectrum results in the strong interband dipole transitions, the polarization sensitive absorption and valley Hall effect~\cite{Yao2008,Matthes2013a,Saroka2022}. Such effects are more pronounced at sizable band gaps and have been observed in monolayers of transition metal dichalcogenides~\cite{Cao2012}. However, they must persist to the optically narrow band gaps, too. Another emergent effect is the matrix element singularity at the Dirac point arising for a decreasing band gap~\cite{Ando2002,Ando2005,Portnoi2009}. Such singularity, presenting a sharp dependence of the dipole transition matrix element on the electron quasi-momentum $\vec{k}$, can manifest itself in a range of graphene-based systems~\cite{Jiang2004,Saroka2017} paving the way to a predictable, well-defined engineering of optoelectronic systems in far-infrared and near terahertz (THz) frequency domain. Its spectroscopic features have been detected in a few subtle experiments~\cite{Itkis2002,Kampfrath2008}.   


Excitons are bound states of an electron and hole, whereby these two particles form a hydrogen atom like structure. Such states can significantly alter the shape of the optical absorption, pushing the absorption edge below the electronic band gap, changing shape of the absorption line, and introducing Sommerfeld factor at the band gap value~\cite{BookHaug2003}. According to Elliot's model excitonic absorption accounts for both the dipole transition matrix element and electron hole probability meet up~\cite{Elliott1957}. Excitonic effects are sensitive to the surrounding environment, screening of the interparticle Coulomb interaction, and system's dimensionality. The latter results in pronounced binding energy increase~\cite{Loudon1959,Ogawa1991,Ogawa1991a}. It has been hypothesized that if the binding energy overcomes the band gap size, an exotic many-body state called an excitonic insulator must form~\cite{Jerome1967}. Many studies, however, predicted that binding energy even in low-dimensional systems scales together with the band gap and never exceeds the latter~\cite{Bulashevich2003,Hartmann2011,Downing2014a,Hartmann2017,Varsano2020,Pulci2023}. Thus, the precise excitonic effects for the low-dimensional systems in the narrow band gap limit is still under debate and careful investigation.

Here, we aim to scrutinize optical properties of a monolayer graphene membrane altered by selective hydrogenation to open a narrow band gap~\cite{Chernozatonskii2007a,Chernozatonskii2008,Chernozatonskii2010}. Such structures can be termed armchair graphene superlattices (AGSLs). They have been demonstrated on graphite by covering the surface with a self-assembled molecular monolayer of cyanuric acid and exposing it to atomic hydrogen~\cite{Nilsson_linear_2012}. Alternative approach could covalent selective patterning~\cite{Navarro2016}. The primary advantage of such systems with respect to quasi-one dimensional tubes and ribbons is their monolithic structure favorable to device engineering. However, the latter requires a theoretical justification due to the strong dependence of the electronic and optical properties on dimensionality, nanostructure geometry, and differences in nanoscale confinement of charge carries in the nanostructures.
The first principles studies of transport in such structures have already demonstrated charge careers channeling along hydrogenation free areas and possibility of imprinting nano-circuits in graphene by controlled hydrogenation~\cite{achilli_hydrogen-dimer_2014}, which gives us promise regarding the optical properties too.
This paper is organized as follows: we present the structures in Sec.~\ref{sec:Structures}, then we consider the optical properties in Sec.~\ref{sec:OpticalProperties}, and finally we summarize our results in Sec.~\ref{sec:Conclusions}. 

\section{\label{sec:Structures}Structural and electronic properties}
\begin{figure*}[t]
\includegraphics[width=0.88\textwidth]{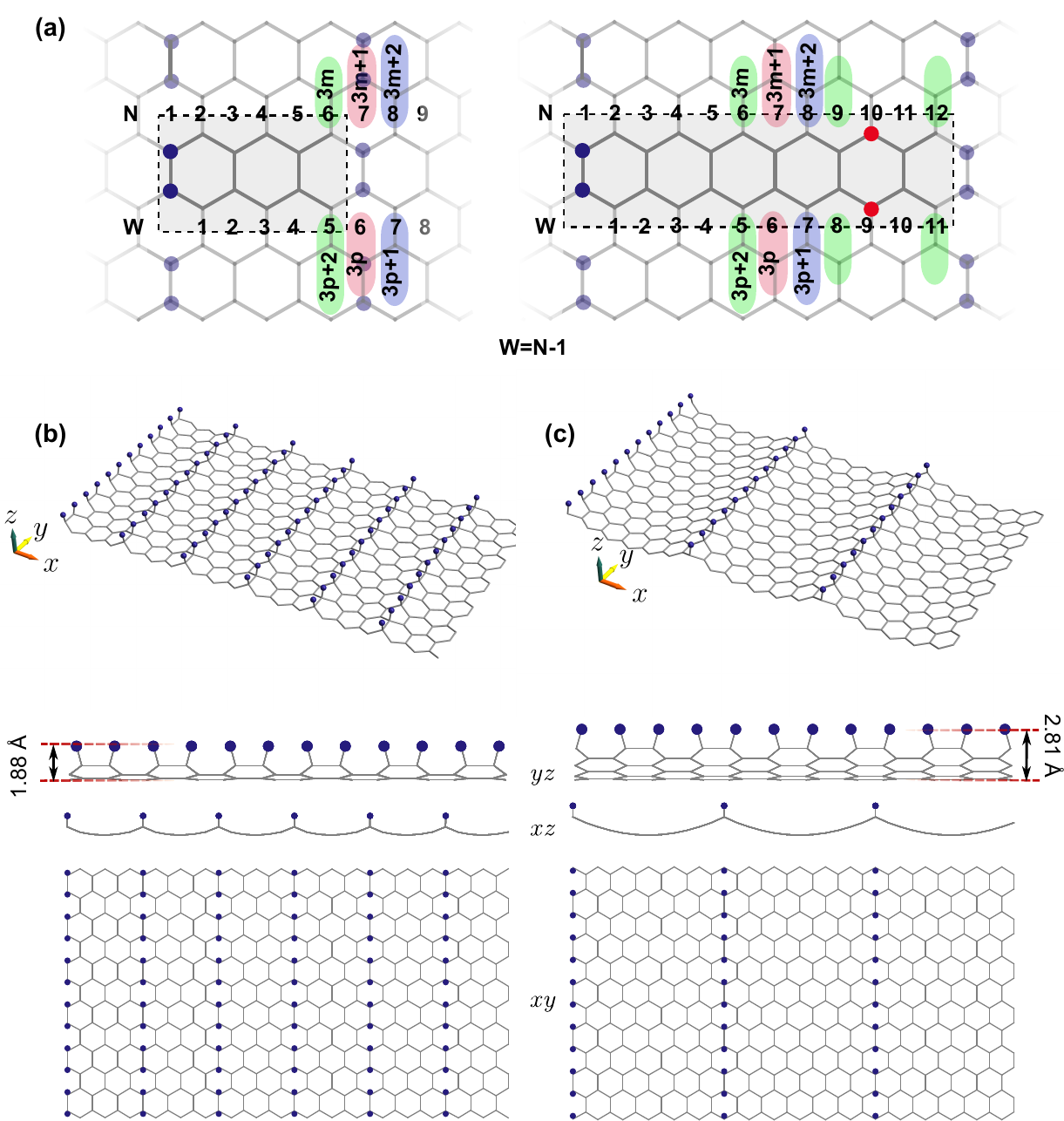}
\caption{\label{fig:AGSLStructure} Structural properties of AGSLs. (a) The schematics of AGSL classification in terms of $N$-index from Ref.~\cite{Chernozatonskii2007a,Chernozatonskii2010} and our $W$-index, exemplified for $W=5$ [left] and $W=11$[right] structures. Hydrogen atoms are dark blue, carbon atoms are omitted for clarity. Dashed rectangles denote unit cells. The semitransparent elements show extension of the unit cells to the 2D lattice. Red disks denote positions of hydrogen atoms that would lead to a different class of superlattice. Light green, red and blue backgrounds highlight the indexes for the three families of the superlattices distinguished by their band gap sizes. (b) The atomic structure of the optimized AGSL($5$). (c) Same as (b) but for optimized AGSL($11$).}
\end{figure*}

In what follows, we use a slightly different notation compared to Refs.~\cite{Chernozatonskii2007a,Chernozatonskii2010}. In Figure \ref{fig:AGSLStructure}, the main structural features and the corresponding notations are presented explicitly. In contrast to Refs.~\cite{Chernozatonskii2007a,Chernozatonskii2010}, where the structures are labeled by the number of carbon atom pairs in the unit cell of the system, $N$, we characterize the width of the superlattices ($Ox$-dimension of the unit cell) by the number of carbon atom pairs $W$, which excludes the carbon pairs covered by hydrogen in the unit cell of the superlattice. As shown in Fig.~\ref{fig:AGSLStructure}(a), this difference introduces the following relation between $N$ and $W$-indexes: $W=N-1$. Thus, to determine the index of the structure, one should take all carbon atoms in the unit cell and divide this number by $2$, which gives the $N$ index. Then the $W$-index is $N-1$. Alternatively, one can count all carbon atoms that are not covered by hydrogen in the unit cell and divide this number by $2$ to obtain $W$ directly. The $W$-index characterizes the number of pairs of carbon atoms squeezed in-between the hydrogen lines. Hereafter we adhere to the following formal notation: AGSL($W$) is the graphene superlattice with armchair crystallographic orientation of adsorbed hydrogen lines separated by $W$ pairs of carbon atoms.

Similarly to zigzag carbon nanotubes~\cite{Dresselhaus1992a,Ajiki1993,Mintmire1993,Yorikawa1994,Ajiki1996a,Kane1997} and armchair graphene nanoribbons~\cite{Nakada1996,Son2006a,Barone2006,Saroka2014a,Saroka2015,Saroka2016a}, hydrogenated superlattices exhibit oscillations of the energy gap with a period defined by the multiplicity of $3$ with respect to the number of carbon atom pairs in the unit cell of the structure. In other words, there are three families of the structures with characteristic band gaps. Since we are interested only in narrow-band gaps, we stick to the family corresponding to the smallest band gap. As can be seen in Fig.~\ref{fig:AGSLStructure}(a), not every formal member of the family can form a unit cell  compatible with that of AGSL($5$). For example, the next formal member requires a different positioning of the hydrogen atoms (red) that is incompatible with the periodicity along the width of the unit cell. The next compatible configuration, however, has a twice larger unit cell corresponding to AGSL($11$). This structural peculiarity explains our choice of these two examples, AGSL($5$) and AGSL($11$), for the model study.

In Figure~\ref{fig:AGSLStructure}(b) and (c), we present the optimized geometries of the free standing membranes corresponding to AGSL($5$) and AGSL($11$), respectively. It is seen that the hydrogenation leads to the $sp^3$-hybridization and a distortion of the flat $sp^2$-hybridized graphene structure around the hydrogenation sites. The larger distance between hydrogen lines results in a larger vertical spread of the membrane, $1.79$~{\AA} versus $2.81$~{\AA}, which provides for such structures more freedom with respect to stretching deformation, which has been shown to enable fine tuning of the band gaps~\cite{Chernozatonskii2010}.
Here, atomic coordinate optimization is performed with the plane-wave based \emph{Quantum Espresso} density functional theory (DFT) code~\cite{Giannozzi2009} using fully nonlocal two-projector norm-conserving Vanderbilt-Hamann pseudopotentials~\cite{Hamann2013} for core electrons with the following convergence criteria: energy $< 10^{-4}$~Ry, forces $<  10^{-3}$~Ry/Bohr, pressure $< 0.5 $~kbar. The valence electron wavefunctions are modeled as linear combinations of plane waves with the kinetic energy cutoff $70$~Ry, discretized on a uniform $k$-grid $10 \times 10$ in the first Brillouin zone. The vacuum separation of about $12$~{\AA} is used between the periodic replicas of the superlattices in the vertical $Oz$-direction.

\begin{figure*}[t]
\includegraphics[width=0.85\textwidth]{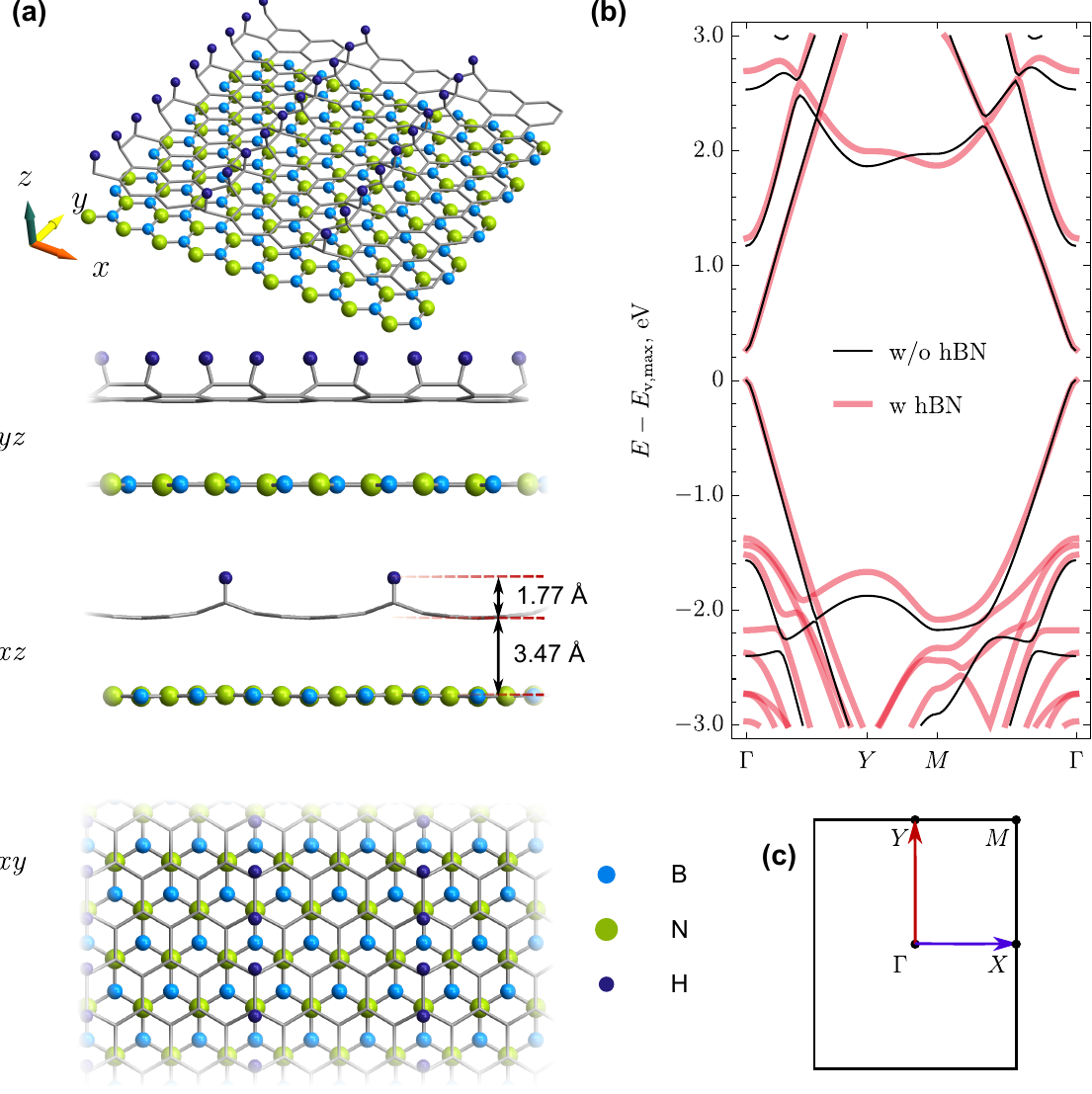}
\caption{\label{fig:AGSLvsAGSLonhBN} The effect of hBN substrate. (a) The optimized structures of AGSL($5$) and AGSL($5$) on hBN substate. (b) The DFT level electronic band structures of AGSL($5$) (w/o) and AGSL($5$) on hBN substate (w). (c) The scheme of Brillouin zone for a rectangular unit cell of AGSL with high symmetry points marked. The scheme also shows half-length reciprocal vectors, $\tfrac{\pi}{T_x}\, \vec{x}$ (blue) and $\tfrac{\pi}{T_y}\, \vec{y}$ (red), where $\vec{x}$ and $\vec{y}$ are unit vectors in reciprocal space.}
\end{figure*}

The free-standing membrane presented above is a model abstraction in a sense. A more realistic system would include a substrate, which, in turn, can alter the structural and electronic properties of a pristine membrane. We note, however, that an atomically smooth and weakly coupled wide band gap dielectric material, in principle, could reinforce the membrane without altering its low-energy physics. The best candidate for this is hexagonal boron nitride (hBN)~\cite{Dean2010a,Xue2011,Tang2013,Mishchenko2014,Mishra2016,Arjmandi-Tash2018,Onodera2020}. Our first principles calculations using Grimme's D3 correction~\cite{Grimme2010} and projector-augmented-wave (PAW) pseudopotentials~\cite{Ahmed2011} support this assumption as follows from Fig.~\ref{fig:AGSLvsAGSLonhBN}. Although the AGSL($5$) membrane has flattened from $1.88$ to $1.77$~{\AA}  due to the introduction of the hBN substrate, the band gap has remained around $0.27$~eV. Therefore, to rationally use computing resources, we proceed with a pristine membrane. 

\section{\label{sec:OpticalProperties}Optical properties}

Although DFT is useful for the initial estimation of electronic and optical properties, it is known to underestimate the material band gaps and to provide only qualitative information on absorption spectra within the independent-particle approximation~\cite{Bechstedt:2015:Book}. For this reason, we performed many-body perturbation theory (MBPT) calculations within the GW approximation for the electronic self-energy~\cite{Hedin1965}, followed by the solution of the Bethe--Salpeter equation (BSE) to obtain an accurate description of the excitonic optical response, as implemented in the YAMBO code~\cite{Sangalli2019}.

Particular attention has been devoted to the convergence of the MBPT calculations, which is especially critical in low-dimensional systems due to the long-range nature of the Coulomb interaction and the strong $\mathbf{q}\rightarrow0$ dependence of the dielectric screening. To address these issues, we employed a computational protocol specifically optimized for 2D materials. In particular, the convergence with respect to the Brillouin-zone sampling was significantly accelerated by combining an interpolation scheme for the screened interaction with a stochastic Monte Carlo integration of both the bare and screened Coulomb potentials, following the approach proposed in Ref.~\cite{Guandalini2023}. This strategy enables an accurate treatment of the long-wavelength screening using substantially denser effective samplings at a reduced computational cost, leading to reliable quasiparticle corrections and excitonic spectra for 2D systems. In absorption spectra, we set the broadening to $0.02$~eV ($20$~meV or $0.2$~ps), which is similar to other works~\cite{Spataru2004,Spataru2004a,Duarte2024}. Note that the excitonic lines in other carbon nanostructures, such as carbon nanostructures, are quite narrow and have larger intrinsic lifetimes of about $10$~ps~\cite{Spataru2005}, but due to energy resolution limitations for the narrow gap structures and 2D nature of the system in question, we use larger values, which can be attributed to other effects such as spatial inhomogenities and phonons.
Detailed convergence tests for all the relevant parameters are described in the reposiory~\cite{DAllesandroGitHub}. We found that for AGSL($5$) the absorption spectra converge using a $20 \times 40$ $k$-point grid and $600$ bands, while for AGSL($11$) convergence is achieved with a $12 \times 36$ $k$-point grid and $600$ bands. This aspect is especially important in light of several reports on excitonic insulating phases in 2D materials~\cite{Jiang2018,Jiang2019,Varsano2020,Sethi2021,Tang2025}, which appear to contradict numerically predicted scaling laws for the exciton binding energy~\cite{Jiang2017a,Pulci2023}. Similar scaling behaviors have also been reported for one-dimensional systems~\cite{Bulashevich2003,Hartmann2011,Downing2014a,Hartmann2017}; therefore, situations in which the exciton binding energy exceeds the quasiparticle band gap may originate from insufficiently converged computational parameters rather than from genuine physical effects \cite{Pulci2023}.

\begin{figure*}[t]
\includegraphics[width=0.85\textwidth]{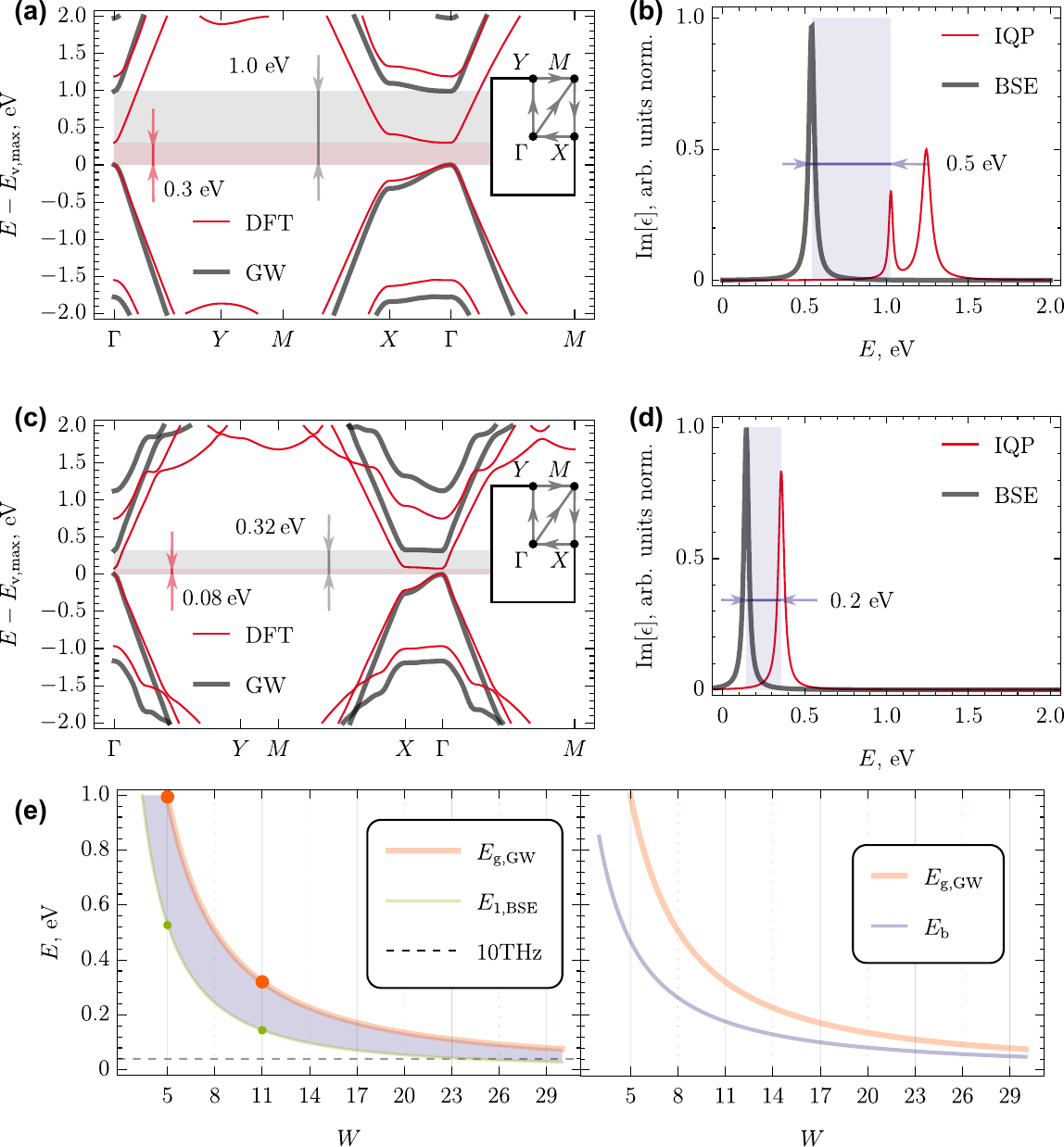}
\caption{\label{fig:TowadsTHz} Towards THz excitons. (a) The electronic band structure of AGSL($5$) at DFT and GW levels. The DFT and GW gaps are highlighted. Inset: $k$-path in the Brillouin zone. (b) The optical absorption spectra at the BSE level for AGSL($5$) compared to the independent quasiparticle (IQP) spectrum. Broadening is $\sim 0.02$~eV. The blueshift of IQP first absorption peak compared to BSE one is highlighted with light blue coloring. (c) Same as (a) but for AGSL($11$). (d) Same as (b) but for AGSL($11$) with broadening $\sim 0.02$~eV. (e) Interpolations for the GW band gaps, $E_{\mathrm{g,GW}}$  (thick, light orange), the first absorption excitation, $E_{1,\mathrm{BSE}}$  (thin, light green), the first exciton binding energy, $E_{\mathrm{b}}$  (light blue), and their extrapolations towards wider AGSL($W$). Solid green and dotted red vertical lines mark the allowed and forbidden members of $3p+2$ family, respectively. The numerically obtained values are shown as dots.}
\end{figure*}

Figure~\ref{fig:TowadsTHz}(a,b) summarizes the electronic and optical properties of AGSL($5$), which is characterized by the close position of the adsorbed hydrogen lines. It is seen from Fig.~\ref{fig:TowadsTHz}(a) that the structure has a direct band gap that is favorable for optical applications. This gap, located at the $\Gamma$ point in the center of the Brillouin zone, is much smaller at the DFT level ($0.3$~eV), compared to the GW calculation ($1.0$~eV). Yet, it is obvious that the shape of the bands does not change that much and the GW bands can be seen as the rigid shift by a fixed quantity of about $0.7$~eV. It is interesting to note that the band configuration between the $X$ and $\Gamma$ points of AGSL is reminiscent of that between $J$ and $K$ points for the diamond surface $(111)$~\cite{Marsili2005,Marsili2008}; this can be attributed to the mixture of the $sp^3$ and $sp^2$ hybridizations in our structure. Concurrently, the bands along the $\Gamma-Y$ high symmetry line, corresponding to the electron momentum parallel to the adsorbed hydrogen lines, show a clear signature of the Dirac cone, which is also characteristic for quasi-metallic armchair nanoribbons and zigzag carbon nanotubes. The latter is a strong evidence of the topological singularity that survives hydrogen adsorption on the surface of graphene. As follows from Fig.~\ref{fig:TowadsTHz}(b), optical transitions across the GW gap are allowed, because this gap is aligned with the edge of absorption in the independent quasi-particle approximation. The excitonic effects, however, drastically change the shape, position, and intensity of the absorption peak. 
Notably, after BSE treatment the peak has a symmetric Lorentzian shape rather than an asymmetric one that could stem from the density of states at the edges of the conduction and valence bands. The first BSE peak is also redshifted to $0.5$~eV due to exciton binding energy and has a prominent intensity, and is not interferred by any other peak upto visible range $2$-$3$~eV. Thus, small AGSL($5$) system corroborates our base hypothesis. The $0.5$~eV BSE peak is far from the THz range; therefore, the next step could be to push this peak to lower energy (frequencies) to achieve THz performance. An obvious approach to achieve this is to increase the separation between hydrogen lines. In this way, the step for momentum quantization could be decreased, and a band gap reduction could be attained. To verify this approach, we study a wider AGSL. 

Figure~\ref{fig:TowadsTHz}(c,d) shows the electronic and optical properties of AGSL($11$). The GW gap for this twice wider structure is $0.32$~eV versus $0.08$~eV in the bare DFT calculation. Other conclusions made with respect to the electronic properties of AGSL($5$) remain valid for AGSL($11$). The optical absorption spectra show a strong absorption peak attributed to the interband transition across the reduced GW band gap, so that the peaks are now shifted to lower energy. The exciton binding energy increases this shift, but the binding energy remains below the GW energy gap ($0.20$ versus $0.32$~eV), ruling out excitonic insulator formation. Thus, we have managed to reduce the energy of the absorption peak by a factor of $\sim 4.4$, approaching the upper border of the THz frequency range, while preserving the primary optical features of interest. 

Since further increasing the width of AGSLs poses challenge for numerical treatment at the high theoretical level, we extrapolate our results in Fig.~\ref{fig:TowadsTHz}, accounting for the fact that in the limit of infinitely wide AGSL the energy band gap must be vanishing. This allows us to fit the GW gaps with the two parameter functional: $a W^{-b}$~\cite{Son2006a,Barone2006,Saroka2014a} with W as defined in Fig.~\ref{fig:AGSLStructure}. The coefficients are found to be $a=10$~eV and $b=1.44$. Similar interpolation is performed for the first absorption peaks in excitonic BSE spectra, where the corresponding parameters are $a=7.43$~eV and $b=1.64$. Given this functional dependence, we estimate that the $10$~THz frequency absorption  ($0.04$~eV) must be achieved at $W=29$ (adjusted to the proper AGSL family, i.e. $3p+2$). It is obvious that further positioned lines of adsorbed hydrogen should be handled more readily because they are well-separated spatially; therefore, we reckon the idea of reaching THz performance with AGSLs to be practical. Figure~\ref{fig:TowadsTHz}(e) also shows the scaling of the exciton binding energy extracted from our numerical data that exclude the excitonic insulator regime at least up to $W=30$.

The variety of carbon nanostructures is expanding rapidly. Many of them are proposed for THz applications. In recent years, carbyne chains~\cite{Kutrovskaya2020,Kucherik2024,Ng2025} and cyclo[n]carbons~\cite{Kaiser2019,Gao2025,Roncevic2026,Grillo2025_circle} have entered this race~\cite{Hartmann2021,Ng2022}. Nevertheless, the membranes considered here possess advantages due to their monolithic 2D structure and numerically corroborated performance in the most ``dense geometry configuration", wherein the hydrogen lines are only one atom thick, while for other nanostructures, assembly into dense arrays may still unpredictably alter their pristine properties~\cite{Polozkov2019} due to Van-der-Waals or even covalent chemical bonding and electromagnetic cross-talk.

We must recognize that observation of THz excitons shall be a very challenging task. This requires dealing with very small spectral broadening and sufficiently clean samples with minimal disorder. In addition, the effects of free charge carriers and doping must be handled with proper care because they can hinder excitonic lines. In fact, to avoid these difficulties the proposed scheme could be first tested in the far infrared regime, where some experiment have been demonstrated~\cite{Ju2017}. After that the excitonic lines could be attempted to push to lower frequencies. We note that free carrier absorption in carbon materials is a broadband effect, that largely depends on the longitudinal size of the system, because it is a geometric antenna resonance~\cite{Shuba2012,Ren2012,Shuba2013}, while the excitonic line strongly depends on the lateral charge carrier confinement as shown above. The latter can be a base for distinquishing the two effects. Likewise, a carrier doping control via backgate and/or suspension~\cite{Sichau2019} could be used for this purpose.

\section{\label{sec:Conclusions}Conclusions}
In summary, we have shown that the optical transitions across a direct band gap of hydrogenated graphene superlattices of armchair type are allowed and strong, thereby leading to a pronounced absorption peak. By varying structural parameters of the lattices, in particular the width of non-hydrogenated strips, it is possible to push this peculiar feature into the THz regime. Notably, the binding energy of such 2D structures, which, in fact, demonstrate quasi-1D-character, is large (upto $\approx 0.5$~eV). The latter facilitates advancing towards THz performance. However, similarly to other carbon nanostructures~\cite{Bulashevich2003,Hartmann2011}, the binding energy in AGSLs scales with the band gap and therefore the excitonic insulator regime seems to be never reachable. Finally, we note that next it would be interesting to investigate stacks of membranes to further increase the material performance. The latter may be possible by replacing hydrogen-induce $sp^3$-hybridization with B-N pairs within graphene.

\begin{acknowledgments}
The authors thank L.~Chernozatonskii for initial assistance with superlattice structures, as well as I.~Kupchak, C.~Hogan, S.~Grillo and D.~Corona for technical support and stimulating discussions. V.A.S. was supported by HORIZON EUROPE MSCA-2021-PF-01 (Project No. 101065500, TeraExc).
\end{acknowledgments}

\bibliography{library}

@PREAMBLE{
 "\providecommand{\noopsort}[1]{}" 
 # "\providecommand{\singleletter}[1]{#1}%" 
}

@book{Bechstedt:2015:Book,
  author = {F. Bechstedt},
  title = {Many-Body Approach to Electronic Excitations. Concepts and Applications},
  publisher = {Springer-Verlag, Berlin},
  year = {2015}
}

@BOOK{BookHaug2003,
   author       = {Haug, Hartmut and Koch, Stephan W.},
   year         = 2003,
   title        = {Quantum Theory of the Optical and Electronic Properties of Semiconductors},
   publisher    = {World Scientific Publishing Co. Pte. Ltd.},
   pagetotal = {465},
}

@ARTICLE{Grillo2025_circle,
	author = {Grillo, Simone and Pulci, Olivia and Giovannini, Tommaso},
	title = {The optical response of aromatic cyclocarbons},
	year = {2025},
    journal = " Chemical Science",
	volume = {16},
	number = {47},
	pages = {22465 – 22472},
	doi = {10.1039/d5sc05519a},
	
}

@ARTICLE{Chernozatonskii2007a,
    author      = "Chernozatonskii, L. A. and Sorokin, P. B. and Belova, E. E. and Br{\"{u}}ning, J. and Fedorov, A. S.",
    title       = "Superlattices consisting of “lines” of adsorbed hydrogen atom pairs on graphene",
    journal     = "JETP Lett.",
    url         = "http://link.springer.com/10.1134/S002136400701016X",
    doi         = "10.1134/S002136400701016X",
    volume      = "85",
    number      = "1",
    pages       = "77--81",
    year        = "2007",
}

@ARTICLE{Chernozatonskii2008,
    author      = "Chernozatonskii, Leonid A. and Sorokin, Pavel B.",
    title       = "Electronic superlattices and waveguides based on graphene: structures, properties and applications",
    journal     = "Phys. Status Solidi B",
    url         = "http://doi.wiley.com/10.1002/pssb.200879578",
    doi         = "10.1002/pssb.200879578",
    volume      = "245",
    number      = "10",
    pages       = "2086--2089",
    year        = "2008",
}

@ARTICLE{Chernozatonskii2010,
    author      = "Chernozatonskii, Leonid A. and Sorokin, Pavel B.",
    title       = "Nanoengineering Structures on Graphene with Adsorbed Hydrogen “Lines”",
    journal     = "J. Phys. Chem. C",
    url         = "http://pubs.acs.org/doi/abs/10.1021/jp9100653",
    doi         = "10.1021/jp9100653",
    volume      = "114",
    number      = "7",
    pages       = "3225--3229",
    year        = "2010",
}

@ARTICLE{Nilsson_linear_2012,
    author      = "Nilsson, Louis and {\v S}ljivan{\v c}anin, {\v Z}eljko and Balog, Richard and Xu, Wei and Linderoth, Trolle R. and L{\ae}gsgaard, Erik and Stensgaard, Ivan and Hammer, Bj{\o}rk and Besenbacher, Flemming and Hornek{\ae}r, Liv",
    title       = "Linear hydrogen adsorbate structures on graphite induced by self-assembled molecular monolayers",
    journal     = "Carbon",
    url         = "https://linkinghub.elsevier.com/retrieve/pii/S0008622311009961",
    doi         = "10.1016/j.carbon.2011.12.050",
    volume      = "50",
    number      = "5",
    pages       = "2052--2056",
    year        = "2012",
}

@ARTICLE{achilli_hydrogen-dimer_2014,
    author      = "Achilli, Simona and Tantardini, Gian Franco and Martinazzo, Rocco",
    title       = "Hydrogen-dimer lines and electron waveguides in graphene",
    journal     = "Phys. Chem. Chem. Phys.",
    url         = "https://xlink.rsc.org/?DOI=C4CP01025F",
    doi         = "10.1039/C4CP01025F",
    volume      = "16",
    number      = "33",
    pages       = "17610--17616",
    year        = "2014",
}

@ARTICLE{Pulci2023,
	author = {Pulci, Olivia and Gori, Paola and Grassano, Davide and D’Alessandro, Marco and Bechstedt, Friedhelm},
	title = {Transitions in Xenes between excitonic, topological and trivial insulator phases: Influence of screening, band dispersion and external electric field},
	year = {2023},
	journal = {SciPost Physics},
	volume = {15},
	number = {1},
	url = {https://www.scopus.com/inward/record.uri?eid=2-s2.0-85167408965&doi=10.21468%2fSciPostPhys.15.1.025&partnerID=40&md5=2786be66f4b2734b3e57101f49e87da5}
}

@ARTICLE{Marsili2005,
    author      = "Marsili, M. and Pulci, O. and Bechstedt, F. and Del Sole, R.",
    title       = "Electronic structure of the {C}(111) surface: {Solution} by self-consistent many-body calculations",
    journal     = "Phys. Rev. B",
    url         = "https://link.aps.org/doi/10.1103/PhysRevB.72.115415",
    doi         = "10.1103/PhysRevB.72.115415",
    volume      = "72",
    number      = "11",
    pages       = "115415",
    year        = "2005",
}

@ARTICLE{Marsili2008,
    author      = "Marsili, Margherita and Pulci, Olivia and Bechstedt, Friedhelm and Del Sole, Rodolfo",
    title       = "Tight-binding calculations of quasiparticle wave functions for {C}(111) 2×1",
    journal     = "Phys. Rev. B",
    url         = "https://link.aps.org/doi/10.1103/PhysRevB.78.205414",
    doi         = "10.1103/PhysRevB.78.205414",
    volume      = "78",
    number      = "20",
    pages       = "205414",
    year        = "2008",
}

@ARTICLE{Hartmann2011,
    author      = "Hartmann, R. R. and Shelykh, I. A. and Portnoi, M. E.",
    title       = "Excitons in narrow-gap carbon nanotubes",
    journal     = "Phys. Rev. B",
    url         = "http://link.aps.org/doi/10.1103/PhysRevB.84.035437",
    doi         = "10.1103/PhysRevB.84.035437",
    volume      = "84",
    number      = "3",
    pages       = "035437",
    year        = "2011",
}

@ARTICLE{Saroka2017,
    author      = "Saroka, V. A. and Shuba, M. V. and Portnoi, M. E.",
    title       = "Optical selection rules of zigzag graphene nanoribbons",
    journal     = "Phys. Rev. B",
    url         = "http://link.aps.org/doi/10.1103/PhysRevB.95.155438",
    doi         = "10.1103/PhysRevB.95.155438",
    volume      = "95",
    number      = "15",
    pages       = "155438",
    year        = "2017",
}

@ARTICLE{Sichau2019,
    author      = "Sichau, J. and Prada, M. and Anlauf, T. and Lyon, T. J. and Bosnjak, B. and Tiemann, L. and Blick, R. H.",
    title       = "Resonance microwave measurements of an intrinsic spin-orbit coupling gap in graphene: A possible indication of a topological state",
    journal     = "Phys. Rev. Lett.",
    url         = "https://doi.org/10.1103/PhysRevLett.122.046403",
    doi         = "10.1103/PhysRevLett.122.046403",
    volume      = "122",
    number      = "04",
    pages       = "046403",
    year        = "2019",
}

@ARTICLE{Duarte2024,
    author      = "Duarte, V. G. M. and Da Costa, D. R. and Peres, N. M. R. and Teles, L. K. and Chaves, A. J.",
    title       = "Moir{\'e} excitons in biased twisted bilayer graphene under pressure",
    journal     = "Phys. Rev. B",
    url         = "https://doi.org/10.1103/PhysRevB.110.035405",
    doi         = "10.1103/PhysRevB.110.035405",
    volume      = "110",
    number      = "03",
    pages       = "035405",
    year        = "2024",
}

@ARTICLE{Novoselov2004,
    author      = "Novoselov, K. S. and Geim, A. K. and Morozov, S. V. and Jiang, D. and Zhang, Y. and Dubonos, S. V. and Grigorieva, I. V. and Firsov, A. A.",
    title       = "Electric field effect in atomically thin carbon films",
    journal     = "Science",
    url         = "http://www.sciencemag.org/cgi/doi/10.1126/science.1102896",
    doi         = "10.1126/science.1102896",
    volume      = "306",
    number      = "5696",
    pages       = "666--669",
    year        = "2004",
}

@ARTICLE{Novoselov2005,
    author      = "Novoselov, K. S. and Geim, A. K. and Morozov, S. V. and Jiang, D. and Katsnelson, M. I. and Grigorieva, I. V. and Dubonos, S. V. and Firsov, A. A.",
    title       = "Two-dimensional gas of massless Dirac fermions in graphene",
    journal     = "Nature",
    url         = "https://doi.org/10.1038/nature04233",
    doi         = "10.1038/nature04233",
    volume      = "438",
    number      = "7065",
    pages       = "197--200",
    year        = "2005",
}

@ARTICLE{Nair2008,
    author      = "Nair, R. R. and Blake, P. and Grigorenko, A. N. and Novoselov, K. S. and Booth, T. J. and Stauber, T. and Peres, N. M. R. and Geim, A. K.",
    title       = "Fine structure constant defines visual transparency of graphene",
    journal     = "Science",
    url         = "http://www.sciencemag.org/cgi/doi/10.1126/science.1156965",
    doi         = "10.1126/science.1156965",
    volume      = "320",
    number      = "1308",
    pages       = "5881",
    year        = "2008",
}

@ARTICLE{Matthes2013a,
    author      = "Matthes, Lars and Pulci, Olivia and Bechstedt, Friedhelm",
    title       = "Massive {Dirac} quasiparticles in the optical absorbance of graphene, silicene, germanene, and tinene",
    journal     = "J. Phys.: Condens. Matter",
    url         = "http://www.ncbi.nlm.nih.gov/pubmed/24002054",
    doi         = "10.1088/0953-8984/25/39/395305",
    volume      = "25",
    number      = "39",
    pages       = "395305",
    year        = "2013",
}

@ARTICLE{Matthes2014,
    author      = "Matthes, L and Pulci, O and Bechstedt, F",
    title       = "Optical properties of two-dimensional honeycomb crystals graphene, silicene, germanene, and tinene from first principles",
    journal     = "New J. Phys.",
    url         = "https://doi.org/10.1088/1367-2630/16/10/105007",
    doi         = "10.1088/1367-2630/16/10/105007",
    volume      = "16",
    number      = "10",
    pages       = "105007",
    year        = "2014",
}

@ARTICLE{Matthes2013,
    author      = "Matthes, Lars and Gori, Paola and Pulci, Olivia and Bechstedt, Friedhelm",
    title       = "Universal infrared absorbance of two-dimensional honeycomb group-{IV} crystals",
    journal     = "Phys. Rev. B",
    url         = "http://link.aps.org/doi/10.1103/PhysRevB.87.035438",
    doi         = "10.1103/PhysRevB.87.035438",
    volume      = "87",
    number      = "03",
    pages       = "035438",
    year        = "2013",
}

@ARTICLE{Fang2013,
    author      = "Fang, Hui and Bechtel, Hans a and Plis, Elena and Martin, Michael C and Krishna, Sanjay and Yablonovitch, Eli and Javey, Ali",
    title       = "Quantum of optical absorption in two-dimensional semiconductors",
    journal     = "PNAS",
    url         = "http://www.pnas.org/content/110/29/11688",
    doi         = "10.1073/pnas.1309563110",
    volume      = "110",
    number      = "29",
    pages       = "11688--11691",
    year        = "2013",
}

@ARTICLE{Merthe2016,
    author      = "Merthe, D. J. and Kresin, V. V.",
    title       = "Transparency of graphene and other direct-gap two-dimensional materials",
    journal     = "Phys. Rev. B",
    url         = "http://link.aps.org/doi/10.1103/PhysRevB.94.205439",
    doi         = "10.1103/PhysRevB.94.205439",
    volume      = "94",
    number      = "20",
    pages       = "205439",
    year        = "2016",
}

@ARTICLE{Tapaszto2008,
    author      = "Tapaszt{\'o}, Levente and Dobrik, Gergely and Lambin, Philippe and Bir{\'o}, L{\'a}szl{\'o} P",
    title       = "Tailoring the atomic structure of graphene nanoribbons by scanning tunnelling microscope lithography",
    journal     = "Nat. Nanotechnol.",
    url         = "https://doi.org/10.1038/nnano.2008.149",
    doi         = "10.1038/nnano.2008.149",
    volume      = "3",
    number      = "7",
    pages       = "397--401",
    year        = "2008",
}

@ARTICLE{Yao2008,
    author      = "Yao, Wang and Xiao, Di and Niu, Qian",
    title       = "Valley-dependent optoelectronics from inversion symmetry breaking",
    journal     = "Phys. Rev. B",
    url         = "https://link.aps.org/doi/10.1103/PhysRevB.77.235406",
    doi         = "10.1103/PhysRevB.77.235406",
    volume      = "77",
    number      = "23",
    pages       = "235406",
    year        = "2008",
}

@ARTICLE{Cao2012,
    author      = "Cao, Ting and Wang, Gang and Han, Wenpeng and Ye, Huiqi and Zhu, Chuanrui and Shi, Junren and Niu, Qian and Tan, Pingheng and Wang, Enge and Liu, Baoli and Feng, Ji",
    title       = "Valley-selective circular dichroism of monolayer molybdenum disulphide",
    journal     = "Nat. Commun.",
    url         = "https://doi.org/10.1038/ncomms1882",
    doi         = "10.1038/ncomms1882",
    volume      = "3",
    number      = "May",
    pages       = "887",
    year        = "2012",
}

@ARTICLE{Jiang2004,
    author      = "Jiang, J. and Saito, R. and Gr{\"u}neis, A. and Dresselhaus, G. and Dresselhaus, M.S.",
    title       = "Optical absorption matrix elements in single-wall carbon nanotubes",
    journal     = "Carbon",
    url         = "https://doi.org/10.1016/j.carbon.2004.07.028",
    doi         = "10.1016/j.carbon.2004.07.028",
    volume      = "42",
    number      = "15",
    pages       = "3169--3176",
    year        = "2004",
}

@ARTICLE{Portnoi2009,
    author      = "Portnoi, M. E. and Rosenau da {Costa}, M. and Kibis, O. V. and Shelykh, I. A.",
    title       = "Magnetically controlled terahertz absorption and emission in carbon nanotubes",
    journal     = "Int. J. Mod. Phys. B",
    url         = "https://doi.org/10.1142/S0217979209062438",
    doi         = "10.1142/S0217979209062438",
    volume      = "23",
    number      = "12",
    pages       = "2846--2850",
    year        = "2009",
}

@ARTICLE{Itkis2002,
    author      = "Itkis, M E and Niyogi, S and Meng, M E and Hamon, M A and Hu, H and Haddon, R C",
    title       = "Spectroscopic study of the {Fermi} level electronic structure of single-walled carbon nanotubes",
    journal     = "Nano Lett.",
    url         = "https://doi.org/10.1021/nl0156639",
    doi         = "10.1021/nl0156639",
    volume      = "2",
    number      = "2",
    pages       = "155--159",
    year        = "2002",
}

@ARTICLE{Kampfrath2008,
    author      = "Kampfrath, T. and von Volkmann, K. and Aguirre, C. M. and Desjardins, P. and Martel, R. and Krenz, M. and Frischkorn, C. and Wolf, M. and Perfetti, L.",
    title       = "Mechanism of the far-infrared absorption of carbon-nanotube films",
    journal     = "Phys. Rev. Lett.",
    url         = "http://link.aps.org/doi/10.1103/PhysRevLett.101.267403",
    doi         = "10.1103/PhysRevLett.101.267403",
    volume      = "101",
    number      = "26",
    pages       = "267403",
    year        = "2008",
}

@ARTICLE{Elliott1957,
    author      = "Elliott, R.",
    title       = "Intensity of optical absorption by excitons",
    journal     = "Phys. Rev.",
    url         = "http://link.aps.org/doi/10.1103/PhysRev.108.1384",
    doi         = "10.1103/PhysRev.108.1384",
    volume      = "108",
    number      = "6",
    pages       = "1384--1389",
    year        = "1957",
}

@ARTICLE{Loudon1959,
    author      = "Loudon, Rodney",
    title       = "One-dimensional hydrogen atom",
    journal     = "Am. J. Phys.",
    url         = "https://doi.org/10.1119/1.1934950",
    doi         = "10.1119/1.1934950",
    volume      = "27",
    number      = "9",
    pages       = "649",
    year        = "1959",
}

@ARTICLE{Ogawa1991,
    author      = "Ogawa, T. and Takagahara, T.",
    title       = "Interband absorption spectra and {Sommerfeld} factors of a one-dimensional electron-hole system",
    journal     = "Phys. Rev. B",
    url         = "http://link.aps.org/doi/10.1103/PhysRevB.43.14325",
    doi         = "10.1103/PhysRevB.43.14325",
    volume      = "43",
    number      = "17",
    pages       = "14325--14328",
    year        = "1991",
}

@ARTICLE{Ogawa1991a,
    author      = "Ogawa, Tetsuo and Takagahara, Toshihide",
    title       = "Optical absorption and {Sommerfeld} factors of one-dimensional semiconductors: An exact treatment of excitonic effects",
    journal     = "Phys. Rev. B",
    url         = "http://link.aps.org/doi/10.1103/PhysRevB.44.8138",
    doi         = "10.1103/PhysRevB.44.8138",
    volume      = "44",
    number      = "15",
    pages       = "8138--8156",
    year        = "1991",
}

@ARTICLE{Jerome1967,
    author      = "J{\'e}rome, D. and Rice, T. M. and Kohn, W.",
    title       = "Excitonic insulator",
    journal     = "Phys. Rev.",
    url         = "https://link.aps.org/doi/10.1103/PhysRev.158.462",
    doi         = "10.1103/PhysRev.158.462",
    volume      = "158",
    number      = "2",
    pages       = "462--475",
    year        = "1967",
}

@ARTICLE{Tang2025,
    author      = "Tang, H. and Yin, L. and Csonka, G. I. and Ruzsinszky, A.",
    title       = "Exploring the exciton insulator state in 1 T-{TiSe}$_2$ monolayer with advanced electronic structure methods",
    journal     = "Phys. Rev. B",
    url         = "https://doi.org/10.1103/PhysRevB.111.L201401",
    doi         = "10.1103/PhysRevB.111.L201401",
    volume      = "111",
    number      = "20",
    pages       = "L201401",
    year        = "2025",
}

@ARTICLE{Spataru2004a,
    author      = "Spataru, C.D. and Ismail-Beigi, S. and Benedict, L.X. and Louie, S.G.",
    title       = "Quasiparticle energies, excitonic effects and optical absorption spectra of small-diameter single-walled carbon nanotubes",
    journal     = "Appl. Phys. A",
    url         = "https://doi.org/10.1007/s00339-003-2464-2",
    doi         = "10.1007/s00339-003-2464-2",
    volume      = "78",
    number      = "8",
    pages       = "1129--1136",
    year        = "2004",
}

@ARTICLE{Spataru2004,
    author      = "Spataru, Catalin D and Ismail-Beigi, Sohrab and Benedict, Lorin X and Louie, Steven G",
    title       = "Excitonic effects and optical spectra of single-walled carbon nanotubes.",
    journal     = "Phys. Rev. Lett.",
    url         = "https://doi.org/10.1103/PhysRevLett.92.077402",
    doi         = "10.1103/PhysRevLett.92.077402",
    volume      = "92",
    number      = "07",
    pages       = "077402",
    year        = "2004",
}

@ARTICLE{Spataru2005,
    author      = "Spataru, C. D. and Ismail-Beigi, S. and Capaz, R. B. and Louie, S. G.",
    title       = "Theory and Ab initio calculation of radiative lifetime of excitons in semiconducting carbon nanotubes",
    journal     = "Phys. Rev. Lett.",
    url         = "https://doi.org/10.1103/PhysRevLett.95.247402",
    doi         = "10.1103/PhysRevLett.95.247402",
    volume      = "95",
    number      = "24",
    pages       = "247402",
    year        = "2005",
}

@ARTICLE{Saroka2022,
    author      = "Saroka, V. A. and Hartmann, R. R. and Portnoi, M. E.",
    title       = "Momentum alignment and the optical valley {Hall} effect in low-dimensional {Dirac} materials",
    journal     = "J. Exp. Theor. Phys. ",
    url         = "http://doi.org/10.1134/S1063776122100107",
    doi         = "10.1134/S1063776122100107",
    volume      = "135",
    number      = "4",
    pages       = "513--530",
    year        = "2022",
}

@ARTICLE{Bulashevich2003,
    author      = "Bulashevich, K. A. and Suris, R. A. and Rotkin, S. V.",
    title       = "Excitons in single-wall carbon nanotubes",
    journal     = "Int. J. Nanosci.",
    url         = "https://doi.org/10.1142/S0219581X03001632",
    doi         = "10.1142/S0219581X03001632",
    volume      = "02",
    number      = "6",
    pages       = "521--526",
    year        = "2003",
}

@ARTICLE{Ju2017,
    author      = "Ju, Long and Wang, Lei and Cao, Ting and Taniguchi, Takashi and Watanabe, Kenji and Louie, Steven G. and Rana, Farhan and Park, Jiwoong and Hone, James and Wang, Feng and {McEuen}, Paul L.",
    title       = "Tunable excitons in bilayer graphene",
    journal     = "Science",
    url         = "https://doi.org/10.1126/science.aam9175",
    doi         = "10.1126/science.aam9175",
    volume      = "358",
    number      = "6365",
    pages       = "907--910",
    year        = "2017",
}

@ARTICLE{Shuba2012,
    author      = "Shuba, M. V. and Paddubskaya, A. G. and Plyushch, A. O. and Kuzhir, P. P. and Slepyan, G. Ya. and Maksimenko, S. A. and Ksenevich, V. K. and Buka, P. and Seliuta, D. and Kasalynas, I. and Macutkevic, J. and Valusis, G. and Thomsen, C. and Lakhtakia, A.",
    title       = "Experimental evidence of localized plasmon resonance in composite materials containing single-wall carbon nanotubes",
    journal     = "Phys. Rev. B",
    url         = "https://doi.org/10.1103/PhysRevB.85.165435",
    doi         = "10.1103/PhysRevB.85.165435",
    volume      = "85",
    number      = "16",
    pages       = "165435",
    year        = "2012",
}

@ARTICLE{Ren2012,
    author      = "Ren, Lei and Pint, Cary L. and Arikawa, Takashi and Takeya, Kei and Kawayama, Iwao and Tonouchi, Masayoshi and Hauge, Robert H. and Kono, Junichiro",
    title       = "Broadband terahertz polarizers with ideal performance based on aligned carbon nanotube stacks",
    journal     = "Nano Lett.",
    url         = "https://doi.org/10.1021/nl203783q",
    doi         = "10.1021/nl203783q",
    volume      = "12",
    number      = "2",
    pages       = "787--790",
    year        = "2012",
}

@ARTICLE{Shuba2013,
    author      = "Shuba, M. V. and Melnikov, A. V. and Paddubskaya, A. G. and Kuzhir, P. P. and Maksimenko, S. A. and Thomsen, C.",
    title       = "Role of finite-size effects in the microwave and subterahertz electromagnetic response of a multiwall carbon-nanotube-based composite: Theory and interpretation of experiments",
    journal     = "Phys. Rev. B",
    url         = "https://doi.org/10.1103/PhysRevB.88.045436",
    doi         = "10.1103/PhysRevB.88.045436",
    volume      = "88",
    number      = "04",
    year        = "2013",
}

@ARTICLE{Withers2011,
    author      = "Withers, F. and Bointon, T. H. and Dubois, M. and Russo, S. and Craciun, M. F.",
    title       = "Nanopatterning of fluorinated graphene by electron beam irradiation",
    journal     = "Nano Lett.",
    url         = "http://pubs.acs.org/doi/abs/10.1021/nl2020697",
    doi         = "10.1021/nl2020697",
    volume      = "11",
    number      = "9",
    pages       = "3912--3916",
    year        = "2011",
}

@ARTICLE{Navarro2016,
    author      = "Navarro, Juan Jes{\'ú}s and Leret, Sof{\'i}a and Calleja, Fabi{\'a}n and Stradi, Daniele and Black, Andr{\'e}s and Bernardo-Gavito, Ram{\'o}n and Garnica, Manuela and Granados, Daniel and V{\'a}zquez De Parga, Amadeo L. and P{\'e}rez, Emilio M. and Miranda, Rodolfo",
    title       = "Organic covalent patterning of nanostructured graphene with selectivity at the atomic level",
    journal     = "Nano Lett.",
    url         = "https://pubs.acs.org/doi/10.1021/acs.nanolett.5b03928",
    doi         = "10.1021/acs.nanolett.5b03928",
    volume      = "16",
    number      = "1",
    pages       = "355--361",
    year        = "2016",
}

@ARTICLE{Saroka2014a,
    author      = "Saroka, V. A. and Batrakov, K. G. and Chernozatonskii, L. A.",
    title       = "Edge-modified zigzag-shaped graphene nanoribbons: {Structure} and electronic properties",
    journal     = "Phys. Solid State",
    url         = "http://link.springer.com/10.1134/S106378341410028X",
    doi         = "10.1134/S106378341410028X",
    volume      = "56",
    number      = "1",
    pages       = "2135--2145",
    year        = "2014",
}

@ARTICLE{Saroka2015,
    author      = "Saroka, V A and Batrakov, K G and Demin, V A and Chernozatonskii, L A",
    title       = "Band gaps in jagged and straight graphene nanoribbons tunable by an external electric field",
    journal     = "Phys. Condens. Matter",
    url         = "http://doi.org/10.1088/0953-8984/27/14/145305",
    doi         = "10.1088/0953-8984/27/14/145305",
    volume      = "27",
    number      = "14",
    pages       = "145305",
    year        = "2015",
}

@ARTICLE{Saroka2016a,
    author      = "Saroka, V. A. and Batrakov, K. G.",
    title       = "Zigzag-shaped superlattices on the basis of graphene nanoribbons: {Structure} and electronic properties",
    journal     = "Russ. Phys. J.",
    url         = "http://link.springer.com/10.1007/s11182-016-0816-6",
    doi         = "10.1007/s11182-016-0816-6",
    volume      = "59",
    number      = "5",
    pages       = "633--639",
    year        = "2016",
}

@ARTICLE{Barone2006,
    author      = "Barone, Ver{\o'}nica and Hod, Oded and Scuseria, Gustavo E",
    title       = "Electronic structure and stability of semiconducting graphene nanoribbons",
    journal     = "Nano Lett.",
    url         = "http://doi.org/10.1021/nl0617033",
    doi         = "10.1021/nl0617033",
    volume      = "6",
    number      = "12",
    pages       = "2748--2754",
    year        = "2006",
}

@ARTICLE{Ng2025,
    author      = "Ng, Raymond A. and Kucherik, Alexey and Portnoi, Mikhail E. and Hartmann, Richard R.",
    title       = "Stark effect tunable terahertz transitions in finite carbon chains",
    journal     = "Small Struct.",
    url         = "https://onlinelibrary.wiley.com/doi/10.1002/sstr.202500266",
    doi         = "10.1002/sstr.202500266",
    volume      = "6",
    number      = "12",
    pages       = "e202500266",
    year        = "2025",
}

@ARTICLE{Downing2014a,
    author      = "Downing, C. A. and Portnoi, M E",
    title       = "One-dimensional {Coulomb} problem in {Dirac} materials",
    journal     = "Phys. Rev. A",
    url         = "http://link.aps.org/doi/10.1103/PhysRevA.90.052116",
    doi         = "10.1103/PhysRevA.90.052116",
    volume      = "90",
    number      = "5",
    pages       = "052116",
    year        = "2014",
}

@MISC{DAllesandroGitHub,
author = {},
title  = {},
year   = {},
url    = {},
note   = {See raw data in GitHub at https://github.com/marcodalessandro76/AGSL5}
}

\end{document}